\documentclass[pra,twocolumn,notitlepage,superscriptaddress,showpacs, amsmath, amssymb]{revtex4-1}

\usepackage[english]{babel}

\usepackage[T1]{fontenc}
\usepackage{textcomp}
\usepackage{amsmath,amssymb}
\usepackage{color, xcolor}
\usepackage{graphicx} 
\DeclareGraphicsExtensions{{.pdf}}
\usepackage{hyperref}
\begin{document}

\title{Ultra-low-loss nanofiber Fabry-P\'erot cavities optimized for cavity quantum electrodynamics}
\author{Samuel K. Ruddell}
\affiliation{Department of Applied Physics, Waseda University, 3-4-1 Okubo, Shinjuku, Tokyo 169-8555, Japan}
\author{Karen E. Webb}
\affiliation{Department of Applied Physics, Waseda University, 3-4-1 Okubo, Shinjuku, Tokyo 169-8555, Japan}
\author{Mitsuyoshi Takahata}
\affiliation{Department of Applied Physics, Waseda University, 3-4-1 Okubo, Shinjuku, Tokyo 169-8555, Japan}
\author{Shinya Kato}
\affiliation{Department of Applied Physics, Waseda University, 3-4-1 Okubo, Shinjuku, Tokyo 169-8555, Japan}
\author{Takao Aoki}
\email{takao@waseda.jp}
\affiliation{Department of Applied Physics, Waseda University, 3-4-1 Okubo, Shinjuku, Tokyo 169-8555, Japan}

\begin{abstract}

\noindent We demonstrate the fabrication of ultra-low-loss, all-fiber Fabry-P\'erot cavities containing a nanofiber section, optimized for cavity quantum electrodynamics. By continuously monitoring the finesse and fiber radius during fabrication of a nanofiber between two fiber Bragg gratings, we are able to precisely evaluate taper transmission as a function of radius. The resulting cavities have an internal round-trip loss of only 0.31\% at a nanofiber waist radius of 207~nm, with a total finesse of 1380, and a maximum expected internal cooperativity of $\sim$~1050 for a cesium atom on the nanofiber surface. Our ability to fabricate such high-finesse nanofiber cavities may open the door for the realization of high-fidelity scalable quantum networks.%
\end{abstract}

\maketitle

\noindent Cavity quantum electrodynamics (CQED) provides a robust platform for the implementation of quantum nodes, which could form the basis of a scalable quantum network~\cite{kimble08, reiserer15}. To maximize the efficiency and fidelity of quantum operations at these nodes, the fabrication of optical cavities having high cooperativity, determined by low loss and high atom--cavity coupling strength, is required~\cite{reiserer15}. It is also necessary to link nodes via a quantum channel with low loss, such that quantum states can be transported between nodes with high efficiency, and entanglement can be distributed across an entire network~\cite{kimble08}. Free-space Fabry-P\'erot cavities have already been used to realize a number of quantum operations, such as deterministic single-photon sources~\cite{mckeever04, hijlkema07}, reversible state-transfer~\cite{boozer07}, quantum gates~\cite{reiserer14, welte18, hacker16}, and nondestructive photon detection~\cite{reiserer13, hosseini16}, as well as the demonstration of an elementary quantum network of two CQED systems connected via a lossy quantum channel~\cite{ritter12}. However, it is technically challenging to implement a large-scale quantum network using such cavities.

To overcome the poor scalability of free-space cavities, fiber-based alternatives present an ideal platform for the realization of large-scale quantum networks. While high finesse fiber-integrated free-space Fabry-P\'erot cavities have been demonstrated~\cite{hunger10, brekenfeld20}, they require precise alignment, and there is intrinsic mode mismatch between the cavity and the fiber. Another candidate is a fiber-coupled whispering-gallery-mode microresonator having ultra-high finesse, where atoms can be coupled via the evanescent field of the cavity~\cite{buck03, spillane05, pollinger09}. While strong coupling to a single atom has been observed~\cite{aoki06, junge13}, and quantum and nonlinear responses at the single-photon level have been demonstrated~\cite{dayan08, aoki09, oshea13, volz14, shomroni14, rosenblum16, bechler18}, microresonators present a number of technical challenges, including trapping of atoms near to the microresonator surface. Silicon nitride nanophotonic cavities have also been demonstrated~\cite{thompson13, tiecke14, samutpraphoot20}, to which atoms can be trapped and coupled using an elaborate optical tweezer technique.

Recently, nanofiber-based cavities have been developed as a promising candidate for CQED~\cite{lekien09, nayak11, wuttke12, schneeweiss17, keloth17}. Due to the tight confinement of the guided mode~\cite{lekien09}, high interaction strengths can be obtained, even with moderate finesse. As these cavities can be directly integrated into a fiber channel, multiple cavity systems can be easily linked. Moreover, nanofiber-based atom traps have been developed~\cite{lekien04, vetsch10, goban12}, and near-ground-state cooling has been demonstrated~\cite{meng18}. Using nanofiber cavities, strong coupling with a single atom in Fabry-P\'erot geometry~\cite{kato15} and collective strong coupling with many atoms in ring cavity geometries~\cite{ruddell17, johnson19} have been observed, as well as dressed states of distant atoms using a network of two coupled cavities~\cite{kato19, white19}. However, as practical nodes in a distributed quantum network, it would be desirable to achieve higher cooperativity by reducing the total loss of the cavity, as well as higher coupling efficiency to the quantum channels by increasing the cavity escape efficiency. The overall performance of the quantum nodes can be characterized by the internal cooperativity~\cite{goto19}, which is the maximal cooperativity in the limit of cavity mirror \mbox{transmission $\rightarrow 0$}, and is inversely proportional to the internal round-trip loss of the cavity and the effective cross-sectional area of the cavity mode.

In this Letter, we present a method for fabricating and optimizing high-finesse fiber Fabry-P\'erot cavities containing a nanofiber section, suitable for use in a CQED-based quantum network. Two fiber Bragg gratings~(FBGs) are written onto an optical fiber by DUV exposure, after which a nanofiber is fabricated between them. By monitoring both the cavity finesse and the nanofiber radius during the tapering process, we are able to optimize the final nanofiber radius to maximize the internal cooperativity of the system by considering the trade-off relation between cavity loss and fiber radius. The finesse is continuously measured using the cavity ringdown technique~\cite{savchenkov07}, while the nanofiber radius is monitored by observing the mode-cutoff times of three lasers with differing wavelengths, transmitted through the cavity during taper fabrication. In this way, we are able to fabricate nanofiber cavities with a radius of 207~nm, having an internal round-trip loss of only 0.31\% and an undercoupled finesse of 1380, measured at a wavelength of $\lambda=852.3$~nm for CQED experiments with cesium atoms. In addition, the FBG reflectivities can be continuously tuned by temperature or tension, allowing the cavity parameters to be easily adjusted. For the cavity presented here, we calculate a maximum internal cooperativity of $\sim1050$ for a cesium atom on the fiber surface.

\begin{figure}[htb]
\begin{center}
\includegraphics[width=\linewidth]{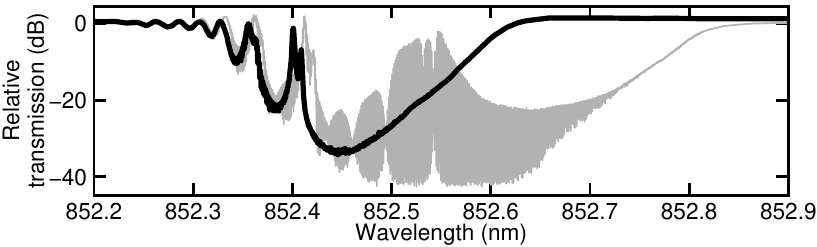}
\end{center}
\vspace{-10pt}
\caption{\fontsize{8}{9.6}\selectfont (a) Measured transmission spectra of a single FBG (black line) and an FBG cavity (gray line). Here the FBGs are tensioned, causing them to be offset from the cesium D$_2$ transition at 852.35~nm.}
\label{fig:fbgtransmission}
\end{figure}
To fabricate an FBG cavity, we first write two FBGs separated by $\sim 24$~cm onto optical fiber (Fibercore SM800) using the UV phasemask method~\cite{hill93}. 
The fiber is stripped and exposed with a 100~mW beam from a 213~nm DUV laser for $\sim60$~minutes. The FBG transmission is monitored during fabrication using a heterodyne setup with $60$~dB of dynamic range. An example of the transmission spectra of a single FBG and an FBG cavity are shown in Fig.~\ref{fig:fbgtransmission}. Utilizing the ringdown technique described later, we then characterize the cavity internal loss by measuring $\mathcal{F}_1$, the finesse of the cavity at critical coupling with light input into one side of the cavity, and $\mathcal{F}_2$, the finesse at critical coupling with light input into the opposite side, tuning only a single FBG to set the coupling between measurements. The total internal round-trip loss is then given by $\alpha_\mathrm{in}=|\pi/\mathcal{F}_1 - \pi/\mathcal{F}_2|$, which we measure to be $\sim0.21$\% for our cavity.

\begin{figure}[htb]
\begin{center}
\includegraphics[width=\linewidth]{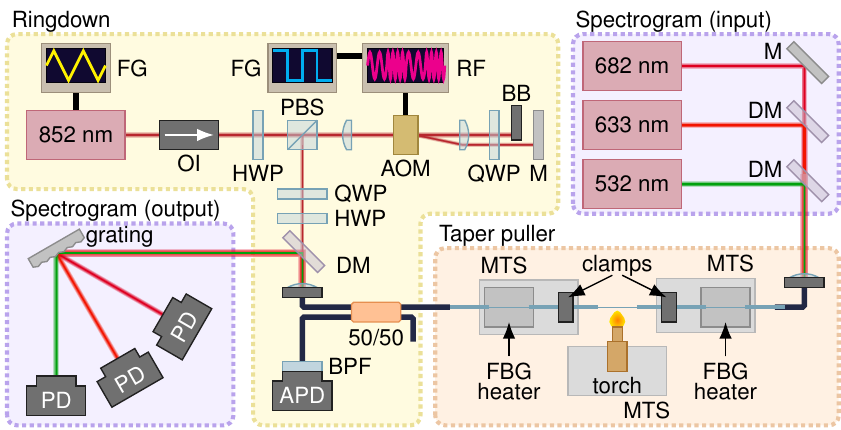}
\end{center}
\vspace{-7pt}
\caption{\fontsize{8}{9.6}\selectfont Experimental setup for the fabrication of optimized nanofiber cavities. FG: function generator; RF: RF function generator; OI: optical isolator; PBS: polarizing beamsplitter; AOM: acousto-optic modulator; HWP: half-wave plate; QWP: quarter-wave plate; BB: beam block; M: mirror; DM: dichroic mirror; PD: photodiode; BPF: bandpass filter; APD: avalanche photodiode; MTS: motorized translation stage.}	
\label{mainsetup}
\end{figure}
The experimental setup used to fabricate and monitor nanofiber cavities consists of three main parts, as shown in Fig.~\ref{mainsetup}. The first is the taper puller used to produce optical nanofibers by the heat and pull method~\cite{birks92, hoffman14}. 
The movement of the stages during tapering is calculated using an optimized taper profile as in~\cite{nagai14}. We choose a nanofiber waist length of 1~mm, compatible with the width of our flame, and an adiabaticity factor of $F=0.3$, as defined in~\cite{nagai14}. With this method, we can reliably produce nanofibers with a radius $\lesssim200$~nm and transmission $\gtrsim99.9$\% for $\lambda=852.3$~nm, with a total taper length of 22~mm.

Secondly, we determine the radius of the nanofiber by measuring the transmission of lasers at different wavelengths passing through the cavity during taper fabrication~\cite{kang20}. By calculating the spectrogram of each transmitted laser, we are able to determine the time of the cutoff point for each higher-order mode and for each wavelength, and hence the corresponding fiber radius at that time~\cite{hoffman14, ravets13}. We use three lasers with wavelengths of 682~nm, 633~nm, and 532~nm, corresponding to single-mode cutoff radii of 248~nm, 228~nm, and 192~nm, respectively, which were chosen to determine the radius of the fiber in the vicinity of 200~nm, close to the optimal nanofiber radius for experiments with cesium atoms~\cite{lekien09}. These lasers are combined using dichroic mirrors to be coupled into the fiber cavity, and detected on three separate photodiodes after being separated by a diffraction grating. As these wavelengths do not fall within the reflection band of the FBGs, they only perform a single pass through the cavity.

The third part of the experimental setup involves the realtime finesse measurement, based on the ringdown technique described in~\cite{savchenkov07}. An external cavity diode laser tuned to 852.3~nm is first sent through an acousto-optic modulator (AOM) in a double-pass configuration, before being coupled into the 50/50 fiber beamsplitter. One output port of the beamsplitter is connected to the cavity, in the opposite direction to the spectrogram lasers, and the light reflected from the cavity is measured at the second input port on an avalanche photodiode (APD). A 10~nm bandpass filter centered at 850~nm, with an extinction ratio $>60$~dB, is placed at the APD input to filter out the $\sim10$~$\mu$W of power from the spectrogram lasers. The AOM is driven by an RF function generator at a center frequency of 110~MHz, and frequency modulated by a 50~kHz square wave with a modulation amplitude of 10~MHz, effectively toggling the laser frequency by 20~MHz every 20~$\mu$s. At the same time, the laser frequency is scanned using a 100~Hz triangle wave in order to slowly sweep through a cavity resonance. Thus, when the laser is resonant with the cavity and the AOM frequency is toggled, a ringdown signal is obtained from the beating between the light exiting the cavity and the now off-resonant laser. 

\begin{figure}[htb]
\begin{center}
\includegraphics[width=\linewidth]{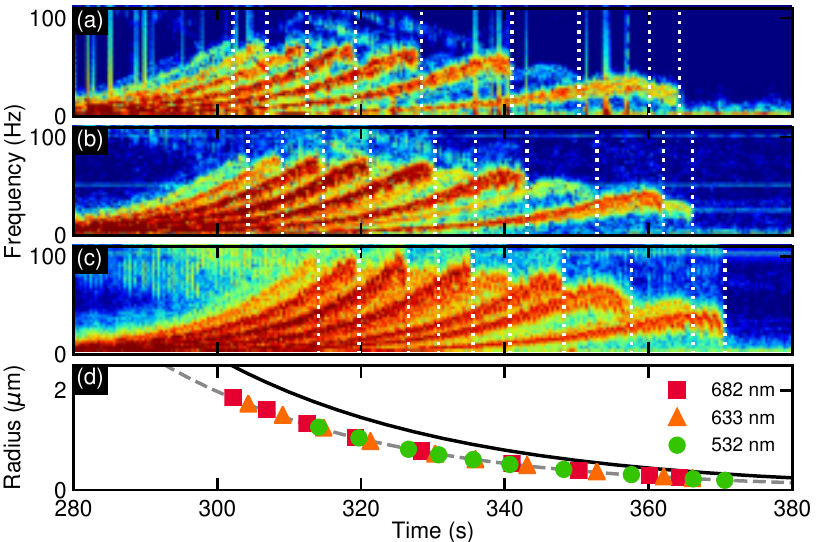}
\end{center}
\vspace{-10pt}
\caption{\fontsize{8}{9.6}\selectfont (a)--(c) Measured spectrograms during the last 100~s of tapering at (a) 682~nm, (b) 633~nm, and (c) 532~nm. Dotted white lines indicate theoretical mode cutoff times. (d) Taper radius as a function of time from the cutoff times in (a)--(c). Solid line shows the theoretical taper profile, while the dashed line shows the adjusted profile accounting for additional lengthening of the fiber due to flame pressure.}
\label{fig:spectrogram}
\end{figure}
Figures~\ref{fig:spectrogram}(a)--(c) show the measured spectrograms in the last 100~s of tapering at 682~nm, 633~nm, and 532~nm, respectively. From the mode cutoff times for each wavelength, we are able to determine the radius as a function of time, as shown in Fig.~\ref{fig:spectrogram}(d). We find that the measured taper profile proceeds slightly faster than the theoretical profile, which we believe is due to pressure exerted onto the fiber from the flame. This effect can be modeled as a small additional lengthening of the fiber during each flame sweep, causing the fiber to become thinner than theoretically expected due to volume conservation. The measured cutoff times agree with this model for an additional lengthening of $6~\mu$m per sweep, shown as the dashed line in Fig.~\ref{fig:spectrogram}(d).

\begin{figure}[htb]
\begin{center}
\includegraphics[width=\linewidth]{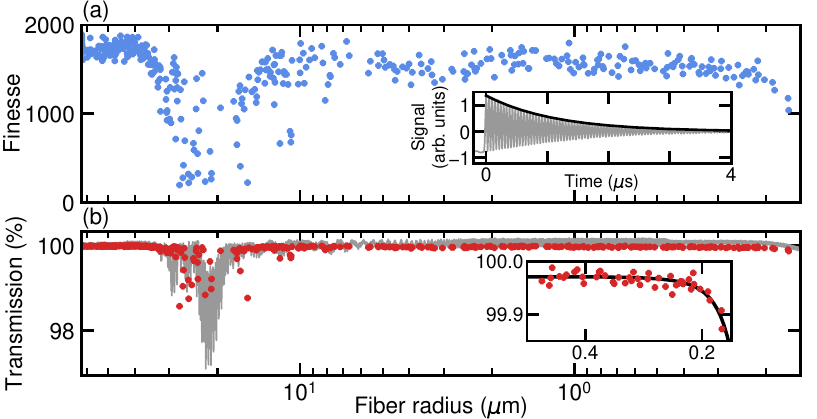}
\end{center}
\vspace{-10pt}
\caption{\fontsize{8}{9.6}\selectfont (a)~Cavity finesse as a function of fiber radius during tapering, where each point represents the finesse from a single ringdown measurement. Inset: example of a single ringdown measurement (gray). Black line shows the exponential fit to the envelope of the ringdown signal. (b)~Taper transmission extracted from the measured finesse in~(a)~\cite{wuttke12} (red circles), and measured single-pass transmission of a taper with the same profile for comparison (gray line). Inset: extracted taper transmission from~(b) near the end of the pull (red circles). Solid line shows an empirical fit to the decrease in taper transmission.}
\label{fig:result}
\end{figure}
Figure~\ref{fig:result}(a) shows the continuous finesse measurement of an FBG cavity during taper fabrication, accounting for the change in free spectral range due to the cavity lengthening, with an example of a single ringdown measurement shown in the inset. For the measurement shown in Fig.~\ref{fig:result}(a), the FBGs were initially tuned to obtain a finesse of $\sim1750$ in the undercoupled regime. As the tapering progresses and the fiber radius becomes thinner, coupling between different spatial modes in the fiber is observed, causing the finesse to fluctuate rapidly. Below a radius of 493~nm, the higher-order symmetric modes of 852~nm light are no longer guided, and these fluctuations cease. The finesse then begins to drop as the fiber radius decreases further, and more light is coupled to radiation modes. From these measurements, we are also able to extract the single-pass taper transmission~\cite{wuttke12}, as shown by the red circles in Fig.~\ref{fig:result}(b). For comparison, Fig.~\ref{fig:result}(b) also shows an example of the taper transmission directly measured when pulling a separate piece of fiber without FBGs (gray solid line). While taper transmission has been conventionally measured in this way~\cite{hoffman14}, it is difficult to precisely measure transmission exceeding 99.9\%, since the transmission must be measured relative to a reference, and the results can be affected by drift of the laser intensity or polarization. In contrast, as the timescale of a ringdown measurement is much faster than the drift of the system, and is independent of the laser intensity, our result represents an absolute measurement of the taper transmission. Indeed, it can be seen from Fig.~\ref{fig:result}(b) that the transmission extracted from the finesse reproduces the overall results from direct measurement, but provides better precision.

In the final stages of pulling, the single-pass taper transmission $T$ as a function of fiber radius $r$ can be modeled as $T(r) = 1-\alpha_0 - (r_0/r)^\tau$, with \mbox{$\alpha_0=0.028$\%}, \mbox{$r_0=49$~nm}, and \mbox{$\tau = 5.8$}, as shown in the inset of Fig.~\ref{fig:result}(b). Here, $\alpha_0$ is the loss due to coupling from the fundamental mode to the higher-order modes arising earlier in the pulling sequence, and $(r_0/r)^\tau$ corresponds to the coupling to radiation modes at $r \lesssim 500$~nm, with $\tau$ determined empirically. By applying the theory of~\cite{sumetsky06, sumetsky06b}, we expect the taper to begin entering the non-waveguiding regime below $r\approx100$~nm, meaning our transmission is not fundamentally limited, and may instead be due to bending of the nanofiber by the pressure exerted from the flame.

Previously, the lowest reported single-pass loss for a nanofiber taper was 0.05\%, measured with 780~nm light for a waist radius of 250~nm~\cite{hoffman14}. To compare, we consider the identical scaled radius $r/\lambda=0.32$, for which we achieve a single-pass loss of only 0.033\%.
The highest finesse nanofiber cavity has been reported in~\cite{keloth17}, where a photonic crystal cavity was imprinted on a 500~nm-diameter nanofiber by femtosecond laser ablation, and therefore the cavity does not include the tapered section, with an estimated round-trip cavity loss of 0.94\%, and a maximum observed finesse of $\mathcal{F}\approx550$ in the undercoupled regime. In contrast, we have achieved an observed undercoupled finesse of $\mathcal{F}\approx1480$ at the same diameter, despite the inclusion of the tapered sections in the cavity.

When fabricating nanofibers for use in CQED systems, there exists an optimum radius where the effective cross-sectional area of the cavity mode is minimized, close to $200$~nm for the cesium D$_2$ line~\cite{lekien09}. As transmission begins to decay below $r\approx 500$~nm, we are able to calculate an optimum nanofiber radius by considering the trade-off relation between radius and internal cavity loss in order to maximize the internal cooperativity \mbox{$C_\mathrm{in} = g^2(\mathbf{r})/2\kappa_\mathrm{in}\gamma$}, which is the maximal cooperativity in the limit of cavity mirror transmission $\rightarrow 0$. Here, $\kappa_\mathrm{in}=c\alpha_\mathrm{in}/(4nL)$ is the cavity field decay rate due only to the round-trip internal cavity loss $\alpha_\mathrm{in}$, with $n$ the refractive index and $L$ the cavity length~\cite{goto19}. $\gamma$ is the polarization decay rate of the atom, and $g(\mathbf{r})$ is the atom--cavity coupling rate given by
$g(\mathbf{r}) = \sqrt{\mu^2\omega/(2\epsilon_0\hbar V_\mathrm{mode})}\phi(\mathbf{r})$,
where $\mu$ is the transition dipole moment, which we take to be the \mbox{$F=4, m_F=0 \rightarrow F^\prime = 5, m_{F^\prime} = 0$} transition of the cesium D$_2$ line, $\phi(\mathbf{r})$ is the cavity mode amplitude, and \mbox{$V_\mathrm{mode} = \int|\phi(\mathbf{r})|^2dV$} is the cavity mode volume~\cite{reiserer15}.

\begin{figure}[htb]
\begin{center}
\includegraphics[width=\linewidth]{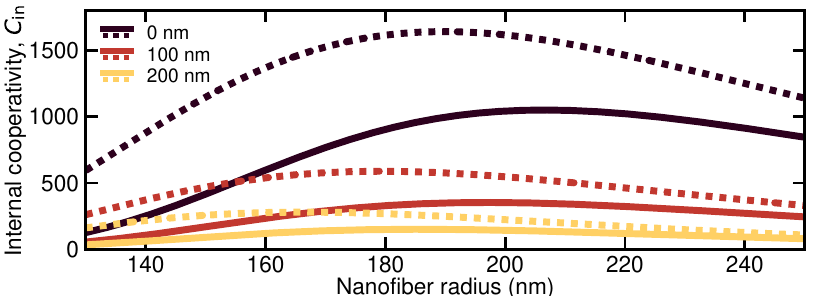}
\end{center}
\vspace{-10pt}
\caption{\fontsize{8}{9.6}\selectfont Calculated internal cooperativity as a function of nanofiber radius for an atom 0, 100, and 200~nm away from the fiber surface. Solid lines include the loss during taper fabrication, while dashed lines indicate the corresponding ideal case of a lossless taper.}
\label{fig:opt}
\end{figure}
In Fig.~\ref{fig:opt}, we show the internal cooperativity as a function of fiber radius, calculated for various atomic distances from the fiber surface, with and without taper loss. For an atom on the surface, we find a maximum \mbox{$C_\mathrm{in}\approx1640$} at a nanofiber radius of 190~nm for the ideal case of no loss due to tapering. However, when we include the taper loss that depends on nanofiber radius, as shown in Fig.~\ref{fig:result}, we find that the maximum \mbox{$C_\mathrm{in}\approx1050$} occurs at a larger nanofiber radius of 207~nm, at which point the single-pass taper transmission is 99.95\%. For an atom trapped 200~nm from the surface, we expect $C_\mathrm{in}\approx150$ at a nanofiber waist radius of 186~nm. For comparison, we calculate the expected $C_\mathrm{in}$ for the cavity in~\cite{keloth17} and find $C_\mathrm{in}=254$ for an atom on the fiber surface, and $C_\mathrm{in}=24$ for an atom trapped 200~nm away.

In summary, we have demonstrated a technique for optimizing the finesse and nanofiber radius of an all-fiber Fabry-P\'{e}rot cavity during nanofiber fabrication. By monitoring both the cavity finesse and the fiber radius, we were able to maximize the expected internal cooperativity of the system. For an atom on the surface, we calculate an internal cooperativity of $C_\mathrm{in}\approx1050$ for a nanofiber radius of 207~nm, with a total round-trip internal cavity loss of only 0.31\%, despite the inclusion of a tapered section within the cavity. These ultra-low-loss cavities could enable the realization of high-efficiency fiber-integrated deterministic single-photon sources, high-fidelity quantum gates and quantum memories in a fiber-based quantum network.

%-----------------------------------------------------------------------------------

\end{document}